# Non-equilibrium thermodynamics of small-scale systems

J.M. Rubi

*Departament de Física Fonamental, Facultat de Física, Universitat de Barcelona, Martí i Franquès, 1, 08028-Barcelona, Spain*

**Abstract**
 Small thermodynamic systems  exhibit peculiar behavior different from that observed in long-scale systems. Non-equilibrium processes taking place in those systems are strongly influenced by the presence of fluctuations which can be large. Contributions to the free energy which vanish at the infinite number of particles limit cannot be neglected and may exert an important influence on the dynamics. We show that in spite of these important differences, the method of non-equilibrium thermodynamics still applies when reducing the size of the system. By using this method, assumption of local equilibrium at the mesoscale thereby leads to the formulation of a mesoscopic nonequilibrium thermodynamics from which expressions for the non-equilibrium currents and kinetic equations for the probability density can be obtained.

Email: mrubi@ub.edu
Fax: 34-93-4021149

## 1. Introduction

Irreversible processes taking place in large-scale systems are well described by non-equilibrium thermodynamics [1]. The theory applies to a coarsened description of the systems which ignores their molecular nature and assumes that they behave as a continuum medium. Under these circumstances, the description does not depend on the size of the system. Scaling up the size does not  lead to new behaviors. Irreversible processes taking place in a small vessel or in an industrial plant can then be analyzed using the same conceptual framework .

When scaling down the size of the system, its molecular nature manifests itself. The coarsening is no longer valid due to the presence of fluctuations in the quantities used in the description. One then reaches a very different scenario in which non-equilibrium thermodynamics may only describe the evolution of the mean values of those quantities but does not completely characterize their actual values. In small systems, such as clusters or biomolecules, otherwise small fluctuations in large-scale systems can be large to the extent that they may become the dominant factor in their evolution. Small is different. The so-called meso-structures, entities whose sizes are in between those of particles and objects sizes, are examples of small systems undergoing assembling, impingement and pattern formation processes in which fluctuations may play a very important role. The knowledge of the functionality of molecular motors, small engines present in many biological systems, and the possibilites of manipulation of matter at small scales to improve performance which constitutes the basic objective of new disciplines as nanoscience and nanotechnology, requires a thermodynamic characterization of the system [2].

Our purpose in this paper is to show that the evolution of small systems can be analyzed by using the method established by non-equilibrium thermodynamics to study irreversible processes taking place at large scales. The new theory proposed to this end, called mesoscopic nonequilibrium thermodynamics [3], can be used to analyze the irreversible behaviour of the meso-structures and to explain processes such as nucleation, growth and active transport in biological membranes which cannot be explained by conventional non-equilibrium thermodynamics [1] due to their inherent nonlinear nature and to the important role played by fluctuations.

## 2. Thermodynamics of small-scale systems

The thermodynamic description of small systems exhibits peculiar features which were already pointed out in the classical book by Hill [2]. Since in those systems the number of particles is not infinite, the free energy may contain contributions not present when the number of particles becomes very large. For example, in a small cluster composed of $N$ particles, the free energy $F$ contains, in addition to the volume term, a surface contribution proportional to $N^{2/3}$. It can be expressed as $F = Nf(T,P) + N^{2/3}g(T,P)$, where $f$ is the free energy per unit of volume and $g$ a function of the temperature $T$ and the pressure $p$. When the number of particles is very large and the system becomes macroscopic size, the surface contribution is negligible and the free energy is simply $Nf$ [2].

The second peculiarity concerns the unavoidable presence of fluctuations at such scales which can be very large, even larger than the mean values. These fluctuations scale with $N^{-1/2}$ and therefore become negligible when the number of particles is very large. These peculiarities become manifest for instance in the study of nucleation phenomena, in the dynamics of single macromolecules and in transport through ionic channels and biological pumps. To formulate the mesoscopic dynamics of thermodynamic systems is then a task of primary importance.

## 3. Mesoscopic non-equilibrium thermodynamics [3]

It is a common feature that the reduction of the observational time and length scales of a system usually entails an increase in the number of nonequilibrated degrees of freedom [4]. Those degrees of freedom, denoted by $\gamma$ ($\equiv \{\gamma_i\}$), may for example represent the velocity of a colloidal particle, the orientation of a magnetic moment, the size of a macromolecule or any coordinate or order parameter whose values define the state of the system in a phase space. The characterization at the mesoscopic level of the state of the system follows from $P(\gamma,t)$, the probability density of finding the system at the state $\gamma \in (\gamma, \gamma + d\gamma)$ at time $t$.

To bring the system to a state characterized by a given value of $\gamma$, we need to exert work. The minimum reversible work required $\Delta W(\gamma)$ is given by

$$\Delta W = \Delta E - T\Delta S + p\Delta V - \mu\Delta M + y\Delta Y + ...,  \quad (1)$$

where $E$ is the internal energy, $S$ the entropy, $V$ the volume, $M$ the mass and $\mu$ the chemical potential. The last term stands for other types of work (electric, magnetic, surface work...) performed on the system, with $y$ being the intensive parameter and $Y$ its conjugated extensive variable [5]. The expression of the work reduces to the different thermodynamic potentials by imposing the constraints that define those potentials. For instance, in the case of a constant temperature, volume and number of particles, the minimum work corresponds to the Helmholtz free energy.

We will assume that the evolution of the degrees of freedom is described by a diffusion process in $\gamma$-space and formulate the corresponding Gibbs equation [6]

$$\delta S = -\frac{1}{T}\int \mu(\gamma)\delta P(\gamma,t)d\gamma, \quad (2)$$

which resembles the corresponding law proposed in nonequilibrium thermodynamics for a diffusion process in terms of the mass density of particles. Here $\mu(\gamma)$ is a generalized chemical potential related to the probability density.

Entropy variations can also be computed from the Gibbs entropy postulate

$$S = S_{eq} - k_B \int P(\gamma,t)\ln\frac{P(\gamma,t)}{P_{eq}(\gamma)}d\gamma, \quad (3)$$

where $S_{eq}$ is the entropy of the system at the equilibrim state in which the probability density is given by

$$P_{eq} \sim \exp\left(\frac{-\Delta W(\gamma)}{k_B T}\right). \quad (4)$$

where $k_B$ is Boltzmann's constant. Taking variations in (3) one obtains

$$\delta S = -k_B \int \delta P(\gamma,t)\ln\frac{P(\gamma,t)}{P_{eq}(\gamma)}d\gamma. \quad (5)$$

where the variations of the equilibrium entropy are given by

$$\delta S_{eq} = -\frac{1}{T}\int \mu_{eq}\delta P(\gamma,t)d\gamma, \tag{6}$$

and $\mu_{eq}$ is the value of the chemical potential at equilibrium. Comparison of both results leads to the identification of the generalized chemical potential

$$\mu(\gamma,t) = k_B T \ln \frac{P(\gamma,t)}{P_{eq}(\gamma)} + \mu_{eq}, \tag{7}$$

which in light of Eq. (4) can also be written

$$\mu(\gamma,t) = k_B T \ln P(\gamma,t) + \Delta W. \tag{8}$$

The evolution in time of the system is then described by a generalized diffusion process over a potential landscape in the space of mesoscopic variables conformed by the values of the equilibrium energy associated to each configuration $\gamma$. The thermodynamic force is $T^{-1}\partial\mu/\partial\gamma$ and the entropy production is given by

$$\sigma = -\frac{1}{T}\int J \frac{\partial \mu}{\partial \gamma} d\gamma \tag{9}$$

or by using the expression of the chemical potential

$$\sigma = -k_B \int J(\gamma,t) \frac{\partial}{\partial \gamma}\left( \ln \frac{P(\gamma,t)}{P_{eq}(\gamma)} \right) d\gamma \tag{10}$$

In this expression, we can then identify the thermodynamic forces as the gradient in the space of mesoscopic variables of the logarithm of the ratio of the probability density to its equilibrium value. The current can be obtained in terms of the force as

$$J(\gamma,t) = -k_B L(\gamma, P(\gamma)) \frac{\partial}{\partial \gamma}\left( \ln \frac{P(\gamma,t)}{P_{eq}(\gamma)} \right), \tag{11}$$

where $L(\gamma, P(\gamma))$ is an Onsager coefficient, which in general depends on the state variable $P(\gamma)$ and on the mesoscopic coordinates $\gamma$. To derive this expression, locality in $\gamma$–space, for which only fluxes and forces with the same value of $\gamma$ become coupled, has also been taken into account.

The resulting kinetic equation follows by substituting Eq. (11) into the continuity equation

$$\frac{\partial P}{\partial t} = -\frac{\partial J}{\partial \gamma}, \tag{12}$$

One then obtains

$$\frac{\partial P}{\partial t} = \frac{\partial}{\partial \gamma}\left(DP_{eq}\frac{\partial}{\partial \gamma}\frac{P}{P_{eq}}\right), \tag{13}$$

where the diffusion coefficient is defined as

$$D(\gamma) \equiv \frac{k_B L(\gamma, P)}{P}. \tag{14}$$

This equation, which in view of Eq. (4) can also be written as

$$\frac{\partial P}{\partial t} = \frac{\partial}{\partial \gamma}\left(D\frac{\partial P}{\partial \gamma} + \frac{D}{k_B T}\frac{\partial \Delta W}{\partial \gamma}P\right), \tag{15}$$

is the Fokker-Planck equation for the evolution of the probability density in $\gamma$-space. Under the conditions for which the minimum work is given by the Gibbs free energy $G$, $\Delta W \equiv \Delta G = \Delta H - T\Delta S$, where $H$ is the enthalpy, this equation transforms into the Fokker-Planck equation for a system in the presence of a free energy barrier:

$$\frac{\partial P}{\partial t} = \frac{\partial}{\partial \gamma}\cdot\left(D\frac{\partial P}{\partial \gamma} + \frac{D}{k_B T}\frac{\partial \Delta G}{\partial \gamma}P\right). \tag{16}$$

Other cases of interest concern different thermodynamic potentials. For instance, a particularly interesting situation is the case of a purely entropic barrier, often encountered in soft condensed matter and biophysics.

We have seen that mesoscopic nonequilibrium thermodynamics provides a simple and direct method to determine the dynamics of a system from its equilibrium properties obtained through the equilibrium probability density.

## 4. Applications

The scheme presented here has been used to analyze non-equilibrium processes of different natures taking place in small-scale systems. In this section, we will mention some cases in which the theory has been successfully applied

*Nucleation kinetics*

The role played by translational and rotational degrees of freedom of the clusters in the nucleation kinetics has been analyzed by using the method of mesoscopic non-equilibrium thermodynamics [7]. It has been shown that the nucleation process is influenced by the dynamics of those degrees of freedom. The expression obtained for the nucleation rate differs from the value of this quantity obtained under the assumption that clusters are at rest. Consideration of mesoscopic variables relaxing in shorter time scales proposes a new scenario which provides a closer agreement with the experiments [8].

*Transport through ionic channels*

Entropic forces due the presence of constraints hampering the access of the system to certain regions play a very important role in many situations such as in the motion of biomolecules through pores, phoretic effects, transport through ionic channels and biological pumps and protein folding. The theory introduced here provides the expression for the diffusion current, the effective diffusion coefficient and the kinetic equation [9]. It has been shown that the diffusion coefficient obeys a scaling law and that the kinetic equation can be formulated under quasi-onedimensional conditions, considerably simplifying the analysis of the diffusion in complex structures.

*Polymer crystallization*

In many practical instances, polymer crystallization takes place under the influence of gradients. The metastable phase under this circumstance has its own nonequilibrium dynamics which is coupled to the crystallization processes. In the case of a velocity gradient [10], the resulting diffusion coefficient of the clusters depends on the shear rate, the pressure and the viscosity and shows anisotropy as found in experiments. These effects become particularly important for high shear rates and small or moderate pressures and when the system is close to the glass transition when the viscosity increases significantly.

*Translocation of a biomolecule*

Many biological processes involve the translocation of proteins or nucleic acids through pores or channels. It has been shown that diffusion drives the biomolecule very slowly in such a way that it would be unable to cross the pore. The presence of proteins binding to the biomolecule rectifies the diffusion and is responsible for the translocation process. The role played by those proteins in the process has been analyzed by considering the number of proteins as a new mesoscopic variable. The value of the resulting force agrees with simulations [11].

*Active transport in biological membranes*

Active transport across membranes is a crucial intermediate step in many biological processes for which ions move against their chemical potential fuelled by the energy released from the hydrolysis of the ATP. Mesoscopic non-equilibrium thermodynmics has been used to derive the nonlinear current for active transport of ions across a membrane [12] consistent with experimental observations. The theory has also been applied to explain the phenomenon of slippage in biological pumps [13].

## 4. Conclusions

We have shown that mesoscopic nonequilibrium thermodynamics describes the irreversible behaviour of small-scale systems in which the presence of large fluctuations exerts an important influence on the dynamics. The theory uses the scheme proposed by nonequilibrium thermodynamics extending the concept of local equilibrium to the mesoscale to show that not only the evolution of the mean values is described by nonequilibrium thermodynamics, as established by Onsager regression laws, but the evolution of the fluctuations as well. The applications to a broad variety of situations [3] and the potential use in others [14] show the usefulness of the theory in providing the mesoscopic dynamics of thermodynamic systems.

## Acknowledgments

We would like to thank S. Kjelstrup, D. Bedeaux, J.M.G. Vilar and D. Reguera for fruitful discussions. This work has been partially supported by the DGICYT of the Spanish Government under Grant No BFM2002-01267.

## References

[1] de Groot SR, Mazur P. Non-equilibrium thermodynamics. New York: Dover, 1984.
[2] Hill TL. Thermodynamics of small systems. New York: Dover, 1994.
[3] Reguera D, Rubi JM and Vilar JMG. The mesoscopic dynamics of themodynamic systems. J. Phys. Chem. B 2005; 109: 21502-15.
[4] Vilar JMG, Rubi JM. Thermodynamics beyond local equilibrium. Proc. Nat. Acad. Sci. 2001; 98: 11081-11084.
[5] Reiss H. Methods of thermodynamics. New York: Dover, 1996.
[6] Pérez-Madrid A, Rubi JM, Mazur P. Brownian motion in the presence of a temperature gradient. Physica A 1994; 212: 231-238.
[7] Reguera D, Rubi, JM. Non-equilibrium translational-rotational effects in nucleation. J. Chem. Phys. 2001 ; 115 : 7100-7106.
[8] Viisanen Y, Strey R, Reiss H.J. Homogeneous nucleation rates for water. J. Chem. Phys. 1993; 99: 4680-4692.
[9] Reguera D, Rubi, JM. Kinetic equations for difusión in the presence of entropic barriers. Phys. Rev. E 2001; 64: 061106(1)-061106(8).
[10] Reguera D, Rubi JM. Nucleation in inhomogeneous media.II. Nucleation in a shear flow. J. Chem. Phys. 2003; 119: 9888-9893.
[11] Zandi R, Reguera D, Rudnick J, Gelbart WD. What drives the translocation of stiff chains? Proc. Nat. Acad. Sci. 2003; 100(15): 8649-8653.
[12] Kjelstrup S, Rubi JM, Bedeaux D. Active transport: a kinetic description based on thermodynamic grounds. J. Theor. Biol. 2005; 234: 7-12.
[13] Kjelstrup S, Rubi JM, Bedeaux D. Energy dissipation in slipping biological pumps. Phys. Chem. Chem. Phys. 2005; 7: 4009-18.
[14] Qian H. Cycle kinetics, steady-state thermodynamics and motors-a paradigm for living-matter physics. J. Phys.: Cond. Mat. 2005; 17; S3783-S3794.